\documentstyle[twocolumn,aps,psfig]{revtex}
\begin{document}
\draft
\twocolumn[\hsize\textwidth\columnwidth\hsize\csname @twocolumnfalse\endcsname
%
%

\title{Magnetic Correlations in the Two Dimensional Anderson-Hubbard Model}

\author{M.~Ulmke$^{1*}$ and R.~T.~Scalettar$^2$}

\address{$^1$Theoretische Physik III, Institut f\"ur Physik,
Universit\"at Augsburg, 
D--86135 Augsburg, Germany
\\
$^2$Department of Physics, University of California, Davis, CA 95616}

\date{\today}
\maketitle

\begin{abstract} 

The two dimensional Hubbard model in the presence of diagonal and
off-diagonal disorder is studied at half filling with a finite 
temperature quantum Monte Carlo method.
Magnetic correlations as well as the electronic compressibility
are calculated to determine the behavior of local magnetic moments,
the stability of antiferromagnetic long range order (AFLRO), and properties 
of the disordered phase. The existence of random potentials (diagonal or 
``site'' disorder) leads to a suppression of local magnetic moments 
which eventually destroys AFLRO. 
Randomness in the hopping elements (off-diagonal disorder), on the 
other hand, does not significantly reduce the density of local magnetic 
moments.  For this type of disorder, at half-filling, there is no
``sign-problem'' in the simulations 
as long as the hopping is restricted between neighbor sites on a
bipartite lattice.
This allows the study of sufficiently large lattices and low temperatures 
to perform a finite-size scaling analysis.  For off-diagonal disorder
AFLRO is eventually destroyed when the fluctuations of antiferromagnetic
exchange couplings exceed a critical value. The disordered phase close to
the transition appears to be incompressible and shows an increase
of the uniform susceptibility at low temperatures.

\end{abstract}

\vskip2pc]
\narrowtext

%
%
\section{Introduction}

Electrons in crystals are scattered both by their mutual interaction
and by static disorder potentials. These processes typically lead to 
quite different or even competing effects. For example, on a bipartite 
lattice close to half-filling the strongly screened Coulomb interaction 
between electrons can generate antiferromagnetic long range order (AFLRO)
while disorder tends to destroy such correlations. The simultaneous 
presence of interaction and disorder cannot in general be considered
as a simple superposition of both contributions but new many-body phenomena
may emerge. This has been found for instance in the study of the 
metal-insulator transitions in doped semiconductors \cite{belitz}
or in the stability of AFLRO against disorder within a dynamical mean-field
theory \cite{ujv}. 
In spite of the great progress that has been achieved in understanding
interacting as well as disordered systems in recent years, there is still 
no controlled and at the same time tractable theoretical method to describe
their combined effects, in particular when the interactions and/or disorder
cannot be considered as small.

It is the purpose of the present paper to provide approximation-free
results for a very simple microscopic model that incorporates electron
interactions as well as disorder, namely the disordered Hubbard model,
or ``Anderson-Hubbard model'' in two dimensions, $D=2$. 
The Hamiltonian of the model reads in the usual notation:
\begin{eqnarray}
\hat H &=&\sum_{<{\bf ij}>,\sigma} t_{\bf ij} \hat c_{{\bf i}\sigma}^\dagger 
        \hat c_{{\bf j}\sigma}
        +\sum_{{\bf i}\sigma} (\epsilon_{\bf i}-\mu) \hat n_{{\bf i}\sigma}
 \nonumber \\
       & & +U\sum_{\bf i}  (\hat n_{{\bf i}\uparrow}-1/2)
                           (\hat n_{{\bf i}\downarrow}-1/2) \; .
\label{model}
\end{eqnarray}

\noindent
$\bf i$ and $\bf j$ are lattice vectors.
The distributions of the random hopping elements $t_{\bf ij}$ and 
random local potentials $\epsilon_{\bf i}$ will be specified later on.
Note that hopping processes are restricted between nearest
neighbors on a square lattice, hence there is no magnetic frustration.

Model (\ref{model}) has been investigated within various approaches: 
the formation of localized
magnetic moments has been studied in $D=3$ for a very broad distribution of 
$t_{\bf ij}$ within an unrestricted Hartree-Fock approximation \cite{sachdev}. 
The case of random potentials has been treated by a real-space renormalization
group method in $D=1,3$ \cite{ma} and $D=2$ \cite{yi}. In $D=2$ this treatment
provides a transition from a Mott to an Anderson insulator with no metallic
phase. However, in these investigations the formation of AFLRO which will set 
in, at least in the unfrustrated case close to half-filling, is not taken
into account. The (in)stability of AFLRO with respect to diagonal disorder 
and two types of metal-insulator transition were examined \cite{ujv} 
in a dynamical mean-field theory which becomes exact in the limit of 
$D\to\infty$ \cite{metzner}. Diagonal disorder has also been studied 
in $D=3$ by Hartree-Fock approximations. \cite{logan} 
The strong-coupling limit of model (\ref{model}) with diagonal 
disorder was studied \cite{zimanyi} using a 
slave-boson formulation  of the corresponding $t-J$ model. 
In the case of off-diagonal disorder at half-filling the model maps in this 
limit onto the spin-1/2  Heisenberg model with random (not-frustrated) 
exchange couplings. 
This model has also been investigated numerically in $D=2$ 
\cite{sandvik2}. 
Finally, the one dimensional Hubbard model with either type of disorder was 
studied using quantum Monte Carlo (QMC) simulations \cite{sandvik1}. 

Here we will concentrate on the effect of disorder on the
magnetic correlations. We will address the following questions:
\begin{itemize}
\item How are short and long range magnetic correlations affected
by the two different kinds of disorder?
\item Is there a critical disorder strength where AFLRO ceases to exist?
\item How do the magnetic susceptibility and charge compressibility
behave in the disordered state?
\end{itemize}

%
%
\section{Computational Method}
We will study the Anderson-Hubbard model in $D=2$ using a finite temperature,
grand canonical quantum Monte Carlo (QMC) method \cite{blankenbecler} which
is stabilized at low temperatures by the use of orthogonalization techniques
\cite{kooninsorella,white}. 
The algorithm is based on a functional-integral representation
of the partition function by discretizing the ``imaginary-time'' interval
$[0,\beta]$ where $\beta$ is the inverse temperature. The interaction
is decoupled by a two-valued Hubbard-Stratonovich transformation \cite{hirsch}
yielding a bilinear time-dependent fermionic action. 
For the positive $U$ model, the $D+1$ dimensional auxiliary field
($s_{{\bf i}l}=\pm 1$ where $\bf i$ is the lattice and $l$ the ``time'' index)
couples to the local magnetization ($n_{i\uparrow} - n_{i\downarrow}$). 
The fermionic degrees of
freedom can be integrated out analytically, and the partition function
(as well as observables) can be written as a sum over the configurations
of the auxiliary field 
with a weight proportional to the product of two determinants, one for 
each spin species. 
The two determinants are not equal since
$s_{{\bf i}l}$ couples with different sign to the two fermion species,
and, in general, their product is not positive definite 
and thus cannot serve as a weight function in an importance sampling procedure.
The formally exact treatment of this ``minus-sign problem'' can lead in some
regimes of the model parameters to very small 
signal-to-noise ratios in physical quantities that become in fact 
exponentially small with inverse temperatures and system size.  

In the case of a bipartite lattice, under the particle-hole 
transformation of one spin species:
\begin{equation}
\hat c_{{\bf i}\downarrow} \longrightarrow 
 (-1)^{\bf i} \hat c_{{\bf i}\downarrow}^\dagger.
\label{phtrafo}
\end{equation}
Hamiltonion (\ref{model}) is mapped onto the $negative$ $U$ Hubbard model:

\begin{equation}
\hat H(U) \longrightarrow \hat H(-U) + \sum_{\bf i} (\epsilon_{\bf i}-\mu) 
(1-2\hat n_{{\bf i} \downarrow}) \; .
\label{negu}
\end{equation}

\noindent
If the Hamiltonion is now spin-symmetric, i.e.~$\epsilon_{\bf i}=\mu$ for
all $\bf i$, the two determinants for spin up and down are identical, since
in the case of the negative $U$ model the Hubbard-Stratonovich field
couples to the charge ($n_{i\uparrow} + n_{i\downarrow}$), that is, 
with the same sign for the two fermion species.
Hence the determinant product is positive (semi) definite and there is no 
``minus-sign problem''.
For this reason the local random potentials $\epsilon_{\bf i}$ lead 
to a minus-sign problem
even at half filling whereas the sign is always positive at half filling
for off-diagonal disorder as long as the hopping remains
restricted between nearest neighbors.

We will consider the case of static, uncorrelated disorder in either the
hopping elements $t_{\bf ij}$ or the on-site potentials $\epsilon_{\bf i}$.
Therefore we have to average all quantities over a sufficient number
of disorder realizations and calculate the averaged expectation values:
\begin{eqnarray}
\langle\langle\hat A \rangle\rangle_h & = & \prod_{<{\bf ij}>} \left[ 
\int dt_{\bf ij} 
P_h(t_{\bf ij}) \right] \; \langle\hat A \rangle (\{t_{\bf ij}\})  \\
\langle\langle\hat A \rangle\rangle_s & = & \prod_{i} \left[ 
\int d\epsilon_{\bf i}
P_s(\epsilon_{\bf i})  \right] \; \langle\hat A \rangle (\{\epsilon_{\bf i}\}).
\label{average}
\end{eqnarray}
$\langle\hat A \rangle$ denotes the thermal expectation value of the operator 
$\hat A$ for a given disorder configuration; $\langle\cdots\rangle_{h(s)}$
stands for the average over hopping (site) disorder.
We will assume uniform distributions of either the hopping elements
$t_{\bf ij}$ with an average value of $t\equiv -1$ or the on-site potentials 
$\epsilon_{\bf i}$ with an average of zero. The width of the distributions 
$P_{h(s)}$ is denoted by $\Delta$:

\begin{eqnarray}
P_h(t_{\bf ij})&=&\frac{1}{\Delta} \Theta ( \frac{\Delta}{2} - |t_{\bf ij}-t|) 
\label{phopping}\\
P_s(\epsilon_{\bf i})&
=&\frac{1}{\Delta}\Theta (\frac{\Delta}{2}-|\epsilon_{\bf i}|) .
\label{psite}
\end{eqnarray}
(We use the same symbol for the width $\Delta$ since each type of disorder
is considered separately.)
Restricting to the half-filled band case $(\mu=0)$ the remaining three
parameters are: interaction $U$, disorder strength $\Delta$, and temperature
$T\equiv 1/\beta$.

In the present study we will concentrate on the following observables:
i) magnetic correlation functions:
\begin{equation}
C({\bf l})=\frac{1}{N} \sum_{{\bf j}} 
\langle\langle \hat m_{\bf j} \hat m_{{\bf j+l}} \rangle\rangle.
\label{correl} 
\end{equation}
Here $\hat m_{\bf j}=\sum_\sigma\sigma \hat n_{{\bf j}\sigma}$ is the local 
spin operator, and $N$ is the total number of lattice sites.
$\sqrt{C(0,0)}$ measures the density of local magnetic moments
and is equal to $n-2d$ with the electronic density 
$n=\sum_{{\bf j}\sigma} \langle\langle\hat n_{{\bf j}\sigma}\rangle\rangle$ 
and the density of doubly occupied sites 
$d=\sum_{\bf j} \langle\langle\hat n_{{\bf j}\uparrow} 
\hat n_{{\bf j}\downarrow}\rangle\rangle$.
(The indices $h,s$ of the disorder averages are suppressed for convenience.) 
ii) magnetic structure factors, the fourier transformation of $C({\bf l})$:
\begin{equation}
S({\bf q}) = \sum_{\bf l} C({\bf l}) e^{i{\bf ql}} \; ,
\label{struct}
\end{equation}
(note that $\beta S(0,0)$ is equal to the uniform spin susceptibility).
iii) charge compressibility:
\begin{equation}
\kappa 
\equiv\frac{\partial n}{\partial\mu}=\frac{\beta}{N}\left[ \sum_{\bf ij} 
\langle\langle\hat n_{\bf i}\hat n_{\bf j} \rangle\rangle \;   -N n^2 \right],
\label{comp}
\end{equation}
with local charge operator 
$\hat n_{\bf j}=\sum_\sigma \hat n_{{\bf j}\sigma}$. 
 
We found $S({\bf q})$ to be largest at the commensurate vector 
${\bf q}=(\pi,\pi)$.
At sufficiently low temperatures the magnetic correlation length exceeds the
system size and $S(\pi,\pi)$ saturates with $\beta$ for a given system size.
Using the saturated values we can extrapolate the behavior in the thermodynamic
limit by a finite size scaling according to spin wave theory \cite{huse} which
predicts in the case of AFLRO in the ground state with sublattice 
magnetization $M$:
\begin{eqnarray}
C(N_x/2,N_x/2)  & = & \frac{M^2}{3} + O(N_x^{-1}) \nonumber \\
\frac{S(\pi,\pi)}{N} & = & \frac{M^2}{3} + O(N_x^{-1}),
\label{scaling}
\end{eqnarray}
where $(N_x/2,N_x/2)$ is the maximal separation on a square lattice of linear
size $N_x=\sqrt{N}$ 
with periodic boundary conditions. Thus, we have two independent 
quantities to extrapolate the value of the ground state order parameter.
The finite size extrapolation is technically only possible in the case 
of off-diagonal disorder where there is no ``minus-sign problem'' and hence
sufficiently large lattices at low temperatures can be simulated.

%
%
\section{Diagonal Disorder}
Diagonal disorder describes the idealized situation of a random alloy 
with negligible lattice distortions but varying values of the chemical 
potentials of the constituents.
While the repulsive interaction $U$ tends to induce singly occupied sites,
i.e.~local magnetic moments, a wide spectrum of random potentials 
has the opposite effect because electrons tend to doubly occupy the 
lower potentials.
Intuitively one would expect that a disorder strength $\Delta$ of the order
of $U$ may be sufficient to destroy AFLRO. This has actually been observed
in the limit $D\to\infty$ \cite{ujv} where the disorder effects are exactly
treated by the ``coherent potential approximation''.\cite{vlaming} 
Since the magnetic moment formation is a local effect we expect 
qualitatively the same behavior in $D=2$.
Fig.~1 shows the local spin-spin correlation fuction, $C(0,0)$, 
on a $4\times 4$ lattice at $U=4$ as a function of $\Delta$. 
$C(0,0)$  decreases monotonically with $\Delta$ and reaches the 
non-interacting value, 0.5, about $\Delta=2U$.
This local effect is indeed independent on dimensionality as it is 
also observed in $D=1$ \cite{sandvik1} and $D=\infty$. \cite{ujv} 
Similar behavior is seen in the spin-spin correlation function at the 
largest separation, $C(2,2)$, and the AF structure factor $S(\pi,\pi)/N$.
They are slightly more stable than  $C(0,0)$ for small disorder and 
decrease more rapidly for $\Delta>4$, also reaching their non-interacting 
values of 0 and 0.0508, respectively about $\Delta=2U$. 

More interesting is the behavior of the compressibility $\kappa$ (Fig.~2).
In a Fermi liquid in the limit $T\to 0$, $\kappa$ is equal to the 
one-particle density of states (DOS) at the Fermi energy. 
For the pure tight binding model without interaction and disorder,
$\kappa$ diverges logarithmically with $T$ at half-filling due to 
perfect nesting.\cite{finitelattices}
Turning on the disorder removes the van-Hove singularity, leading to
a finite DOS at the Fermi level and to a broadening of the DOS.
For large $\Delta$ the DOS is dominated by the disorder spectrum,
giving a bandwidth proportional to $\Delta$, and due to normalization
a value at the Fermi energy proportional to $1/\Delta$. 
This relation is also observed for the compressibility at $U=0$ 
at a finite temperature (Fig.~2). 

\begin{figure}
\vskip-08mm
\psfig{file=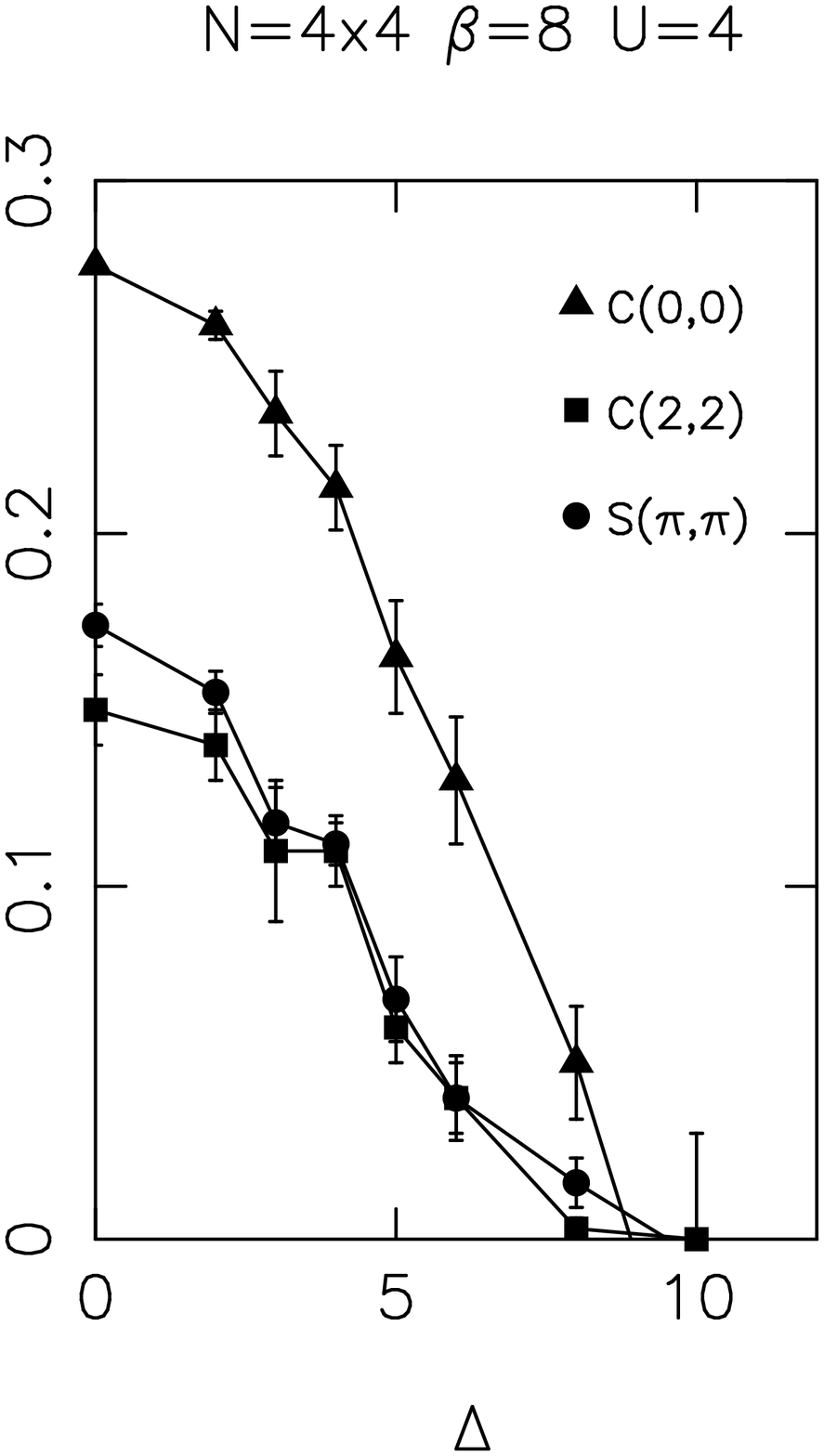,height=3.5in,width=4.in}
\vskip-03mm
\caption{Local moment $C(0,0)$ [triangles], 
longest range spin correlation $C(2,2)$ [squares]
and antiferromagnetic structure factor $S(\pi,\pi)/N$ 
[circles] on 4x4 lattices at $U=4$ and $T=1/8$,
as a function of site disorder $\Delta$.  The noninteracting
values of each quantity have been subtracted out.}
\end{figure}

\begin{figure}
\vskip-10mm
\psfig{file=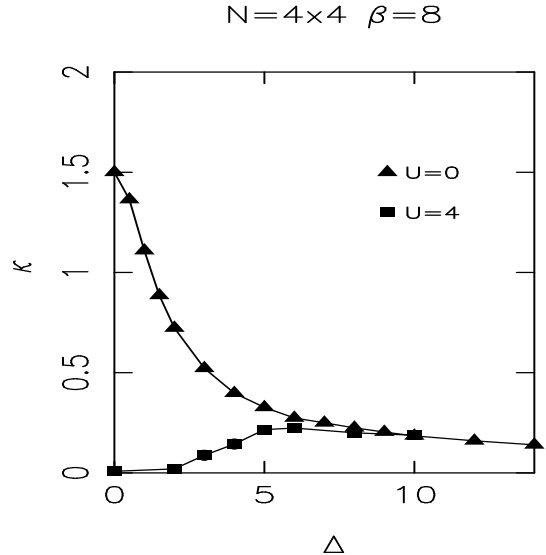,height=3.5in,width=4.in}
\vskip-3mm
\caption{Compressibility $\kappa$ 
at $U=0$ [triangles] and $U=4$ [squares] on 4x4 lattices
and $T=1/8$, as a function of site disorder $\Delta$.
The very different behavior at $\Delta=0$ reflects the
suppression of the divergent density of
states in the noninteracting limit and the opening
of a charge gap.}
\end{figure}

On the other hand, when the interaction strength $U$ is nonzero,
for small values of $\Delta$ 
the compressibility vanishes due to the charge gap induced 
by antiferromagnetic ordering. AFLRO is strictly present only
in the ground state, but for a finite lattice the AF correlation length 
at a finite temperature can exceed the lattice size. \cite{white}
Even with a fully established charge gap, $\kappa$ is always finite
at finite temperature, however exponentially suppressed (Fig.~2).
$\kappa$ starts to increase significantly for $\Delta>2$ and reaches 
a maximum at about $\Delta=6$. Since the $\kappa$ vs.~$\Delta$ curve for
$U=4$ approaches the non-interacting curve for large $\Delta$
the finite compressibility is 
quite likely not thermally activated but due to the closing of the charge gap.
Further, $\kappa$ becomes finite for relatively small values of $\Delta<4$
where the AF correlations are still strong. Although the lattice size
is much too small for a definite conclusion this indicates that the
charge gap may close within the AF phase. Evidence for a
disorder induced AF metal at half filling has previously been found in 
$D=\infty$ \cite{ujv} and also in a Hartree-Fock treatment in $D=3$.
\cite{logan} 
Note, however, that in the presence of disorder a finite compressible does 
not necessarily imply metallicity because of localization effects.

Fig.~3 shows why we cannot study large lattices 
and low temperatures in the case of diagonal disorder.  The averaged
values of the sign of  the product of determinants,  
$\langle\langle \sigma \rangle\rangle$,
vanishes rapidly with spatial lattice size $N$ and inverse temperature
$\beta$.  A value  $\langle\langle \sigma \rangle\rangle$  smaller than about
0.2 precludes reliable simulations due to a vanishing signal to noise ratio
in the data. 

\begin{figure}
\vskip-08mm
\psfig{file=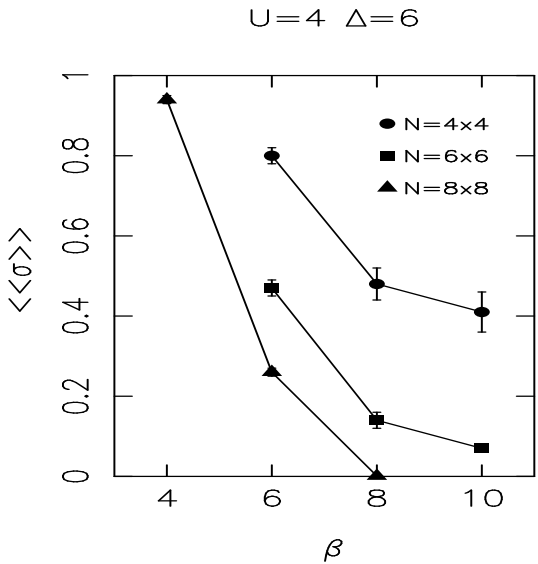,height=3.5in,width=4.in}
\vskip-3mm
\caption{Average sign $\langle \langle \sigma \rangle \rangle$
for the site disordered problem
at $U=4$ and $\Delta=3$ as a function of inverse temperature
$\beta$. Circles are 4x4 lattices, while
squares and triangles are 6x6 and 8x8 lattices respectively.
This sign problem is absent in the case of hopping disorder at half-filling.}
\end{figure}

%
%
\section{Off-diagonal Disorder}

In the case of random hopping elements
$t_{\bf ij}$ restricted to near neighbor sites on a bipartite lattice
there is no ``minus sign problem'' at half filling.
Therefore we can do a much more detailed analysis of the phase diagram.
We study square lattices with periodic boundary conditions
up to the size $N=100$ ($N_x=10$). For a given disorder configuration,  
500-700 Monte Carlo sweeps were performed for equilibration
followed by 1000-1500 measurement sweeps.
Then, all measured quantities were averaged over 10-20 different disorder 
configurations. The disorder average is the main source of the statistical
errors.

As in the previous section, we first study the spin-spin correlations 
as a function of disorder strength $\Delta$ for a given lattice size.
Fig.~4a shows $C(\bf l)$ over a path in real space for different values of 
$\Delta$ on a 6x6 lattice. 
AF correlations are present for all values of $\Delta$.  They are
only slightly
reduced for $\Delta\le 0.8$ and much more significantly reduced 
when $\Delta \sim 1.6$.
Larger lattices show a quite similar picture (Fig.~4b).
The local moments $\sqrt{C(0,0)}$ are apparently stable for off-diagonal
disorder, as is shown in Fig.~5 where the behavior
of $C(\bf l)$ for ${\bf l}=(0,0)$ and $(N_x/2,N_x/2)$ and the
AF structure factor $S(\pi,\pi)/N$  are plotted as a function of $\Delta$.
Unlike the case of random site energies (Fig.~1), $C(0,0)$ is 
almost unchanged by $\Delta$. 
However, measures of the long range
order are strongly affected. While the longer range correlations break
down for strong disorder ($\Delta\sim 2$) they slightly \em increase \em
for small $\Delta$.
This slight increase between $\Delta=0.0$ and $0.2$ might be due
to an enhanced averaged Heisenberg exchange coupling
\begin{equation}
\langle J_{\bf ij} \rangle = \frac{4\langle t_{\bf ij} ^2 \rangle }{U}
= J_0 + \frac{\Delta^2}{3U},
\label{javerage}
\end{equation}
with $J_0=4t^2/U$ for the not disordered case.
However this is essentially a strong coupling argument $(J_0\ll t)$
while the present value is only $J_0=t$.

In order to answer the question if there is AFLRO in the ground state
we calculate $C(N_x/2,N_x/2)$ and $S(\pi,\pi)$ for different lattice
sizes and extrapolate the results assuming a finite size scaling 
according to (\ref{scaling}).
For the present lattice sizes (up to $10\times 10$ sites) 
the magnetic correlations are saturated at a temperature of 
about $T=1/10$ where the finite system is essentially in its 
ground state.  
For disorder strength $\Delta\le 1.2$ both $C(N_x/2,N_x/2)$ and 
$S(\pi,\pi)$ extrapolate to a nonzero order parameter $M$
at $N_x\to\infty$ (Fig.~6).
The values for $M$ are obtained by a least-square fit of the data
with $N_x\ge6$. For $N_x=4$ there are apparently deviations from
the scaling (11).
Within the statistical error the independent extrapolations lead 
to the same value of $M$. 
For $\Delta=1.6$, however, there is no long range order.

\begin{figure}
\vskip-08mm
\psfig{file=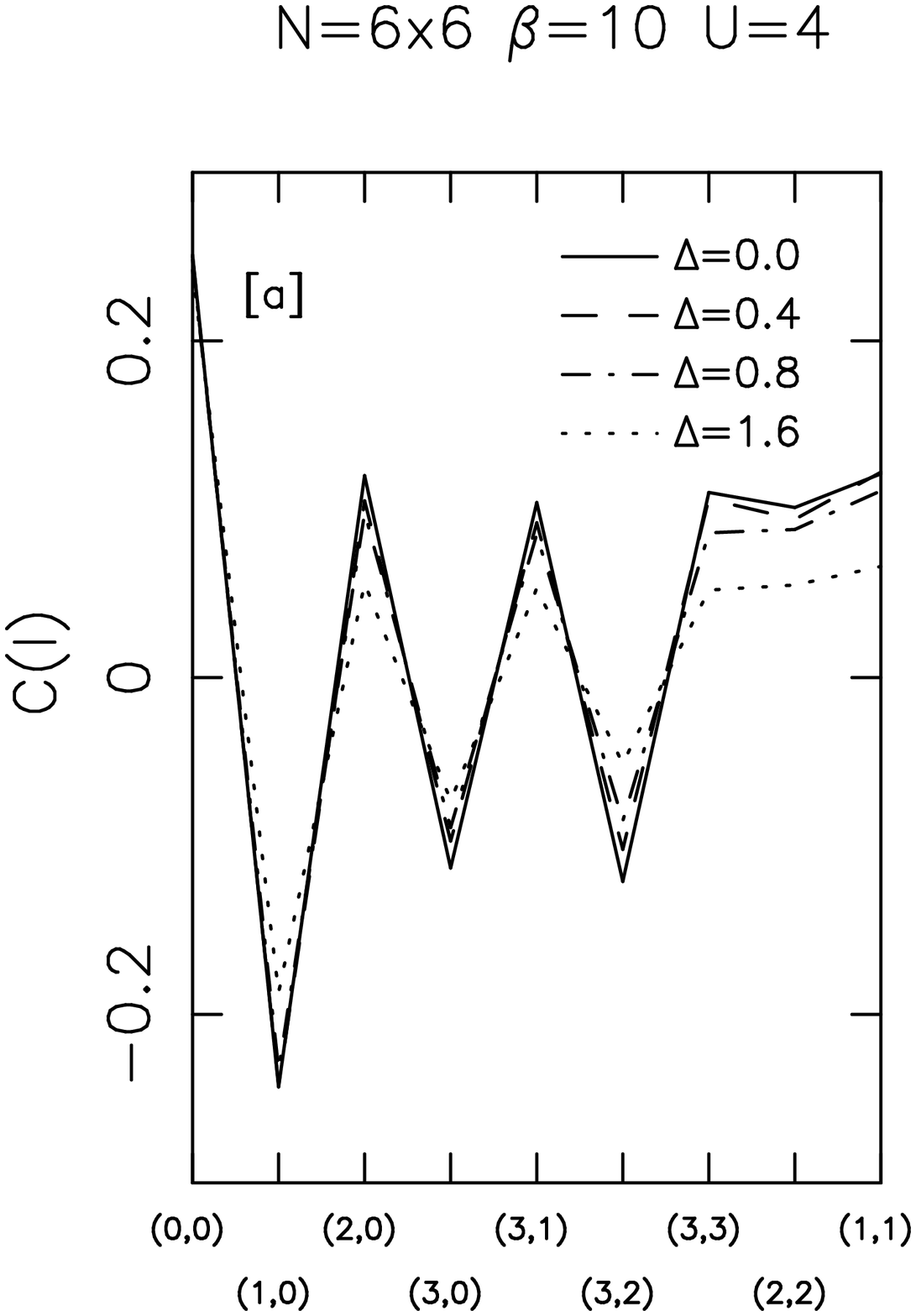,height=3.5in,width=4.in}
\vskip-08mm
\psfig{file=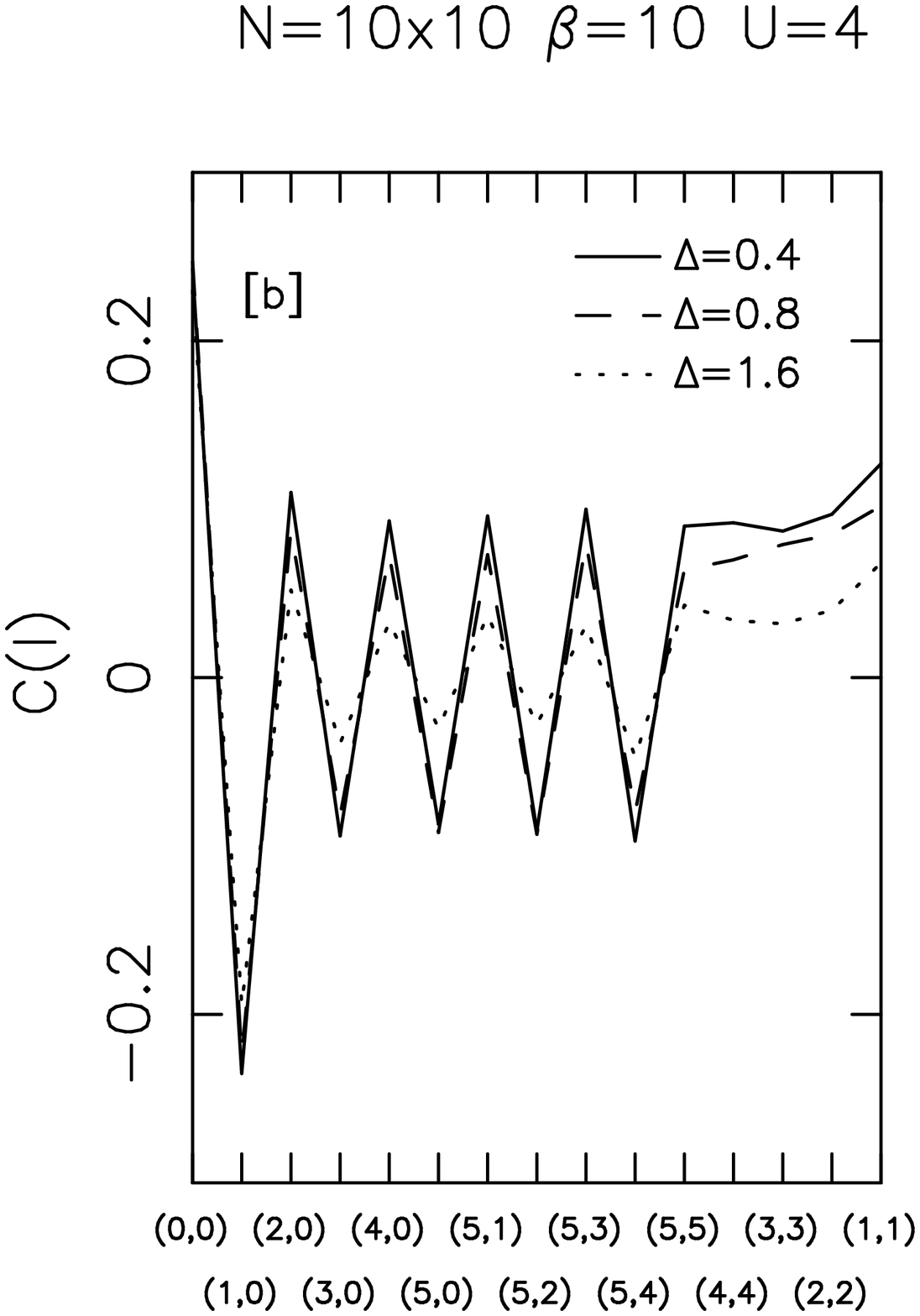,height=3.5in,width=4.in}
\vskip-03mm
\caption{Spin-spin correlations as a function of separation
on (a) 6x6 and (b) 10x10 lattices at $T=1/10$ and $U=4$.
The value shown at $(0,0)$ is $C(0,0)-1/2$.}
\end{figure}

The values of $M$ as a function of $\Delta$ are shown in Fig.~7
where the value for $\Delta=0$ is taken from the literature \cite{white}.
$M$ is apparently stable for small disorder strength $\Delta\le 0.8$
and then decreases and eventually vanishes about $\Delta\approx 1.4$. 
The slight increase of magnetic correlation for small disorder at a 
given system size (Fig.~5) is not observed in the values of the order 
parameter.

\begin{figure}
\vskip-08mm
\psfig{file=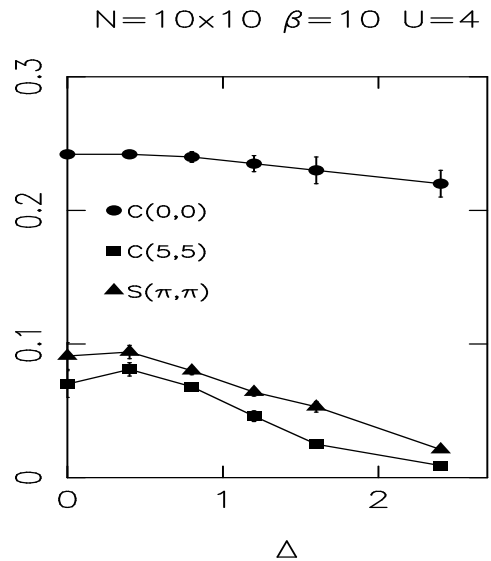,height=3.5in,width=4.in}
\vskip-3mm
\caption{Local moment $C(0,0)$ [circles], 
longest range spin correlation $C(5,5)$ [squares]
and antiferromagnetic structure factor $S(\pi,\pi)$/N 
[triangles] on 10x10 lattices at $U=4$ and $T=1/10$,
as a function of site disorder $\Delta$. The noninteracting
values of each quantity have been subtracted out.}
\end{figure}

The physical reason for the destruction of AFLRO is not obvious
for this type of disorder. There is no magnetic frustration and also
no destruction of local moments.  One possible approach
to understanding the transition is a weak-coupling analysis.
Within the RPA the magnetic susceptibility of the interacting
system $\chi_{RPA}$ is expressed in terms of the noninteracting
value $\chi_{0}$:
\begin{equation}
\chi_{RPA}(\pi,\pi)  =   \frac{\chi_{0}(\pi,\pi)}
{1-U\chi_{0}(\pi,\pi)}.
\label{RPACLEAN}
\end{equation}
The perfect nesting relation of the pure tight binding model
without interactions and disorder: 
$\varepsilon({\bf q}+(\pi,\pi))) = -\varepsilon({\bf q})$ 
at half filling leads to a divergent susceptibility
\begin{equation}
 \chi_0(\pi,\pi)  \propto  (\ln T)^\gamma.
\label{chi0}
\end{equation}
For the square lattice at half filling we have $\gamma=2$ due to the 
logarithmic van-Hove singularity in the DOS at the Fermi energy.
In general, for bipartite lattices without this DOS singularity like the 
simple cubic lattice the exponent becomes $\gamma=1$.
Since $\chi_{0}$ diverges at $T=0$ the RPA predicts that the
system will order at any finite $U$.
    
One effect of the disorder is
removing the perfect nesting relation of the pure tight binding model,
reducing $\chi_{0}$ and leading to the destruction of AFLRO within this
RPA analysis.
However, off-diagonal disorder does not remove the bipartite structure of
the lattice, so we expect Eq.~(\ref{chi0}) still to hold with $\gamma=1$
although $\bf q$ is no longer a good quantum number.
Since there exists no

\begin{figure}
\vskip-08mm
\psfig{file=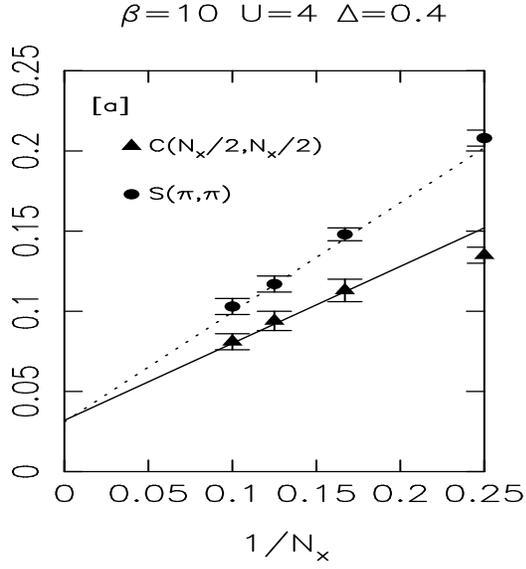,height=3.5in,width=4.in}
\vskip-8mm
\psfig{file=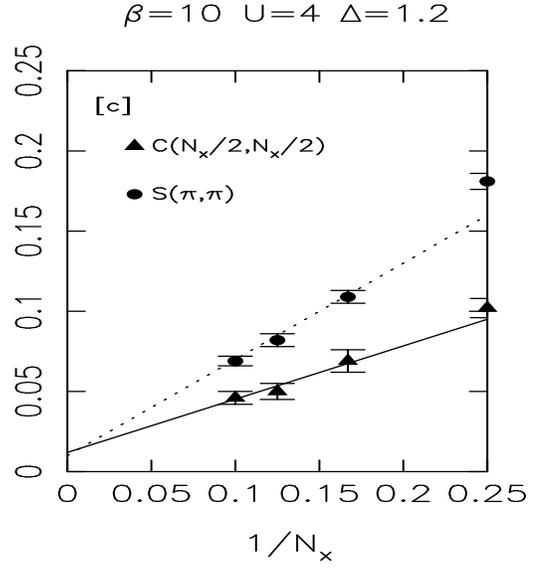,height=3.5in,width=4.in}
\vskip-8mm
\psfig{file=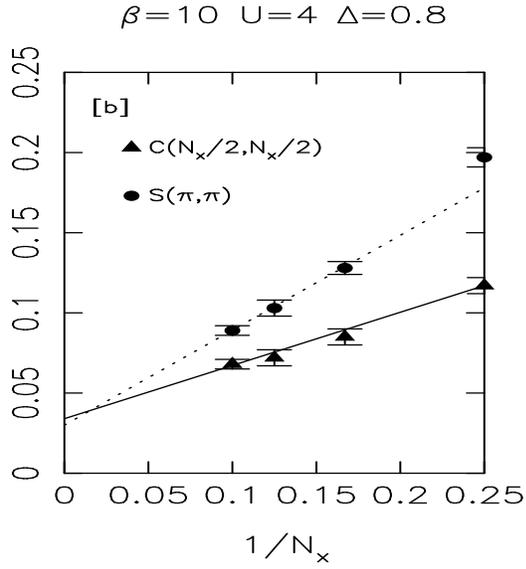,height=3.5in,width=4.in}
\vskip-8mm
\psfig{file=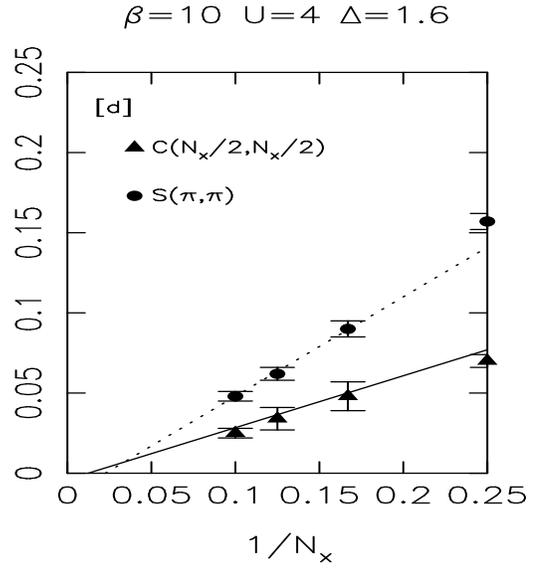,height=3.5in,width=4.in}
\vskip-3mm
\caption{Finite size scaling analysis of the AF
structure factor and long range spin correlations.
These quantities, appropriately scaled, are plotted as a function
of the inverse linear lattice size $1/N_{x}$.  A non-zero
extrapolation to $1/N_{x}=0$ indicates AFLRO.
The extrapolated values should be identical.}
\end{figure}
    
\begin{figure}
\vskip-08mm
\psfig{file=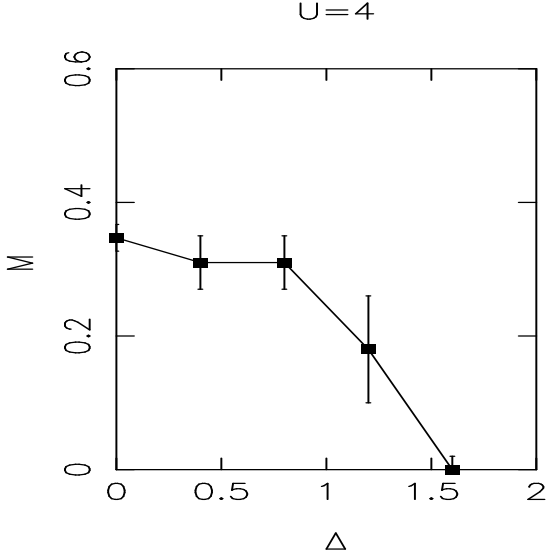,height=3.5in,width=4.in}
\vskip-3mm
\caption{The order parameter $M$ inferred
from Fig.~6 as a function of $\Delta$.
For $\Delta < \Delta_{c} \approx 1.4$ there is AFLRO.  
For $\Delta > \Delta_{c} \approx 1.4$ there is
a transition into a disordered phase.}
\end{figure}

\noindent
analytic solution for $\langle\chi(\bf q) \rangle$ 
for the disordered lattice even without interaction, we calculate it
numerically for finite lattices.
$\chi_0(\pi,\pi)$ is indeed well 
described by a logarithmic fit (Fig.~8).
In fact, we further
observe that in the disorder-averaged RPA susceptibility 
$\chi_{RPA}(\pi,\pi)$
for finite $U$ one can to a very good approximation replace 
$\chi(\pi,\pi)$ for a given disorder configuration by its
average value $\langle \chi(\pi,\pi) \rangle$:
\begin{eqnarray}
\chi_{RPA}(\pi,\pi) & = & \left\langle \frac{\chi(\pi,\pi)}
{1-U\chi(\pi,\pi)}\right\rangle
\approx \nonumber \\
\chi_{approx.}(\pi,\pi) & = & \frac{\langle \chi(\pi,\pi) \rangle}
{1-U\langle \chi(\pi,\pi)\rangle}\; . 
\label{RPA}
\end{eqnarray}
In particular the estimated 
Ne\'el temperatures for given $U$
obtained from the divergence of $\chi_{RPA}(\pi,\pi)$ and 
$\chi_{approx.}(\pi,\pi)$ are identical within the errors due 
to disorder averaging (Fig.~8).
We observe that $\chi_{RPA}({\bf q})$ for different momenta 
${\bf q}\ne (\pi,\pi)$ is always more strongly suppressed by disorder
than $\chi_{RPA}(\pi,\pi)$. 
We conclude that off-diagonal disorder does not remove the
low temperature divergence of the AF susceptibility.
Thus an explanation of the transition based, for example,
on the inducement of a finite
$U_{crit} > 4$ as the disorder is increased is not viable.

While the above considerations might be suited in the case of
weak interaction, we will now consider the opposite case
of $U\gg t$. Here, the half filled Hubbard model with random $t_{\bf ij}$
at half filling becomes equivalent to the disordered, but $unfrustrated$, 
AF spin-1/2 Heisenberg model with random couplings $J_{\bf ij}$, see 
(\ref{javerage}).
This model has been studied in $D=2$ by a Quantum Monte Carlo 

\begin{figure}
\vskip-08mm
\psfig{file=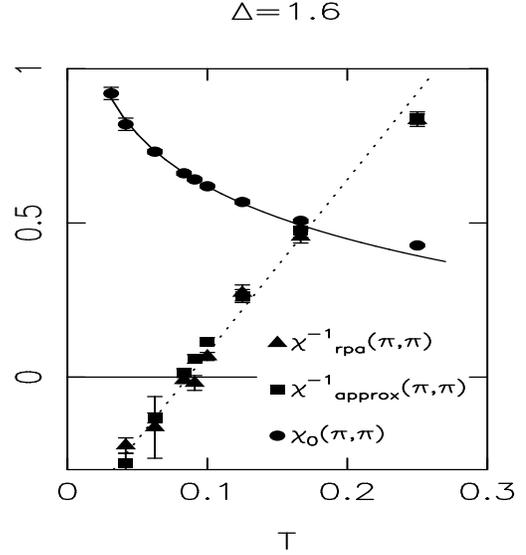,height=3.5in,width=4.in}
\vskip-3mm
\caption{The behavior of the noninteracting susceptibility $\chi_{0}$ (circles)
with temperature $T$ in the presence of disorder.  
$\chi_0$ still diverges (logarithmically) as $T\rightarrow 0$. 
The full line is a fit with $\chi_0=0.054-0.245\ln T$.
Also shown are the inverse RPA susceptibility
 $\chi_{RPA}^{-1}(\pi,\pi)$ (triangles) and the approximation 
$\chi_{approx.}^{-1}(\pi,\pi)$ (squares). 
They are indistinguishable within the statistical errors
and vanish at the same critical temperature. The dotted line
is a guide to the eye only.}
\end{figure}

\noindent
technique in the case of a binary distribution 
of $J_{\bf ij}\in \{J_1,J_2\}$ 
\cite{sandvik2}.  This numerical work incorporated
disorder in two ways-  both through the difference of the exchange energies
$\tilde\Delta=|J_1-J_2|$ and also through 
different concentrations of strong and weak bonds.
Depending on the concentration $p$ of, say the $J_1$ bonds, and the 
difference $\tilde\Delta$, an AF ordered and a disordered phase were 
identified.  The basic physics is that spin singlets
form on the strong bonds, driving 
the formation of a paramagnetic phase, as has
been discussed by Bhatt and Lee.\cite{BHATT,BELITZ1}
We have made a quantitative comparison of our data for the
Hubbard model at $U=4$ with this strong coupling theory which
incorporates spin degrees of freedom only.
We first note that the result of \cite{sandvik2},
for the phase boundary in the $p-\tilde\Delta$ plane
is reasonably well described by setting the variance of 
$J_{\bf ij}$,
\begin{equation}
v \equiv \frac{\langle J_{\bf ij}^2 \rangle - \langle J_{\bf ij}\rangle^2}
 {\langle J_{\bf ij}^2 \rangle}
= \frac{p(1-p)\tilde\Delta^2}{pJ_1^2+(1-p)J_2^2},
\label{variance}
\end{equation}
to a critical value $v=v_c$. \cite{percolation}
That is, the fact that there are two distinct types of randomness appears
irrelevant-  they can be modeled together in a simple way by their 
effect on the variance of $J_{\bf ij}$.
The fit to the calculated phase boundary is best for $v_c\approx 0.40$ to 0.42.
The variance of $J$ in the Hubbard model with disorder distribution 
(\ref{phopping}) is independent of $U$ and reads:
\begin{equation}
v = \frac{\langle t_{\bf ij}^4 \rangle-\langle t_{\bf ij}^2\rangle^2}
 {\langle t_{\bf ij}^4 \rangle} = \frac{4}{9} \left(
\frac{\Delta^4+60t^2\Delta^2}{\Delta^4+40t^2\Delta^2+80} \right),
\end{equation}
$v$ has its maximum of $5/9$ at $\Delta=4\sqrt{5}t$.
If we now apply the same criterion, $v=v_c$, as in the Heisenberg model
we obtain a critical disorder strength of $\Delta_c=1.7t$ to $1.8t$. 
This estimation is surprisingly close to the critical
$\Delta$ obtained by finite size scaling at $U=4$ (Fig.~7) in spite
of the fact that $U=4$, corresponding to $J_0=1$ is not in the strong 
coupling limit of the Hubbard model.\cite{doublelayer}
The agreement might be due to the fact that already for $U=4$
the double occupancies are strongly suppressed and the density of
local moments has reached about 75\% of the maximal value 1.0
independent of disorder strength (see Fig.~5).
Further, $\Delta_c$ is small compared to $3U$ so that the averaged
$\langle J_{\bf ij}\rangle$ is only slightly enhanced (by a factor of 
$\sim 1.25$) according to (\ref{javerage}) which is a consistency check of 
the arguments above. 

To characterize the properties of the disordered phase $(\Delta>1.4)$ 
we study the temperature dependence of the uniform spin susceptibility
$\chi(0,0)=\beta S(0,0)$ (\ref{struct}) and the charge compressibility
$\kappa$ (\ref{comp}). In the non-interacting case both quantities are
identical, and diverge logarithmically in $D=2$ for $\Delta=0$,
as mentioned before. \cite{finitelattices} 
This singularity is removed by interactions, and
$\chi(0,0)$ reaches a maximum at a finite temperature, approaching a finite
value at $T=0$. \cite{white}   
Disorder suppresses the non-interacting susceptibility too,
but for a different reason because it removes the van-Hove singularity 
in the density of states (Fig.~9a). In the interacting case however,
disorder has the opposite effect of enhancing $\chi(0,0)$ if the 
disorder is strong enough $(\Delta=1.6)$. For $\Delta\le 0.8$
there is no significant effect on $\chi(0,0)$ due to disorder at the 
lowest temperature under consideration $(T=t/10)$.
In the present temperature regime at $\Delta=1.6$, $\chi(0,0)$ 
decreases monotonically 
with $T$, a behavior also found in the disordered phase of the random bond
Heisenberg model \cite{sandvik2}.
Stronger disorder ($\Delta=2.4$, see Fig.~9b), where some of the hopping
elements become negative, leads to further enhancement of $\chi(0,0)$.  
In systems with longer range interactions a low temperature divergence
of $\chi(0,0)$ has been predicted. \cite{BHATT} This divergence is explained 
by the presence of localized moments, i.e.~spins on lattice sites which are
effectively decoupled from the rest of the system. Such isolated sites are
particulary likely in the case of a special type of \em correlated \em
hopping randomness with $t_{\bf {ij}}=x_{\bf i} x_{\bf j}$ where the site 
variables $\{x_{\bf i}\}$ are randomly distributed. \cite{kotliar}
In contrast to the uncorrelated distribution (\ref{phopping}), here a site
$\bf i$ with a small variable $x_{\bf i}$ is weakly connected to $all$ its 
neighbors.
We consider a bimodal, symmetric  distribution of site 
variables:
\begin{equation}
P(x_{\bf i})=\frac{1}{2} \left\lbrack 
\delta(x_{\bf i}-\sqrt{t-\frac{\Delta'}{2}})+
\delta(x_{\bf i}-\sqrt{t+\frac{\Delta'}{2}})
\right\rbrack
\end{equation}
with $\Delta'\le 2t$. 
This distribution corresponds to a binary 
alloy where the hopping amplitudes $t_{\bf {ij}}$ depend only 
on the (three possible) combinations of atoms on 
neighboring sites.
$t_{\bf {ij}}$ are always positive and can take the 
values $t-\Delta'/2$, $\sqrt{t^2-(\Delta'/2)^2}$, and $t+\Delta'/2$. 
The average  Heisenberg coupling remains 
$\langle J_{\bf ij}\rangle=J_0=4t^2/U$ independent of $\Delta'$.
We find that for $\Delta'=1.6$, $\chi(0,0)$ 

\begin{figure}
\vskip-08mm
\psfig{file=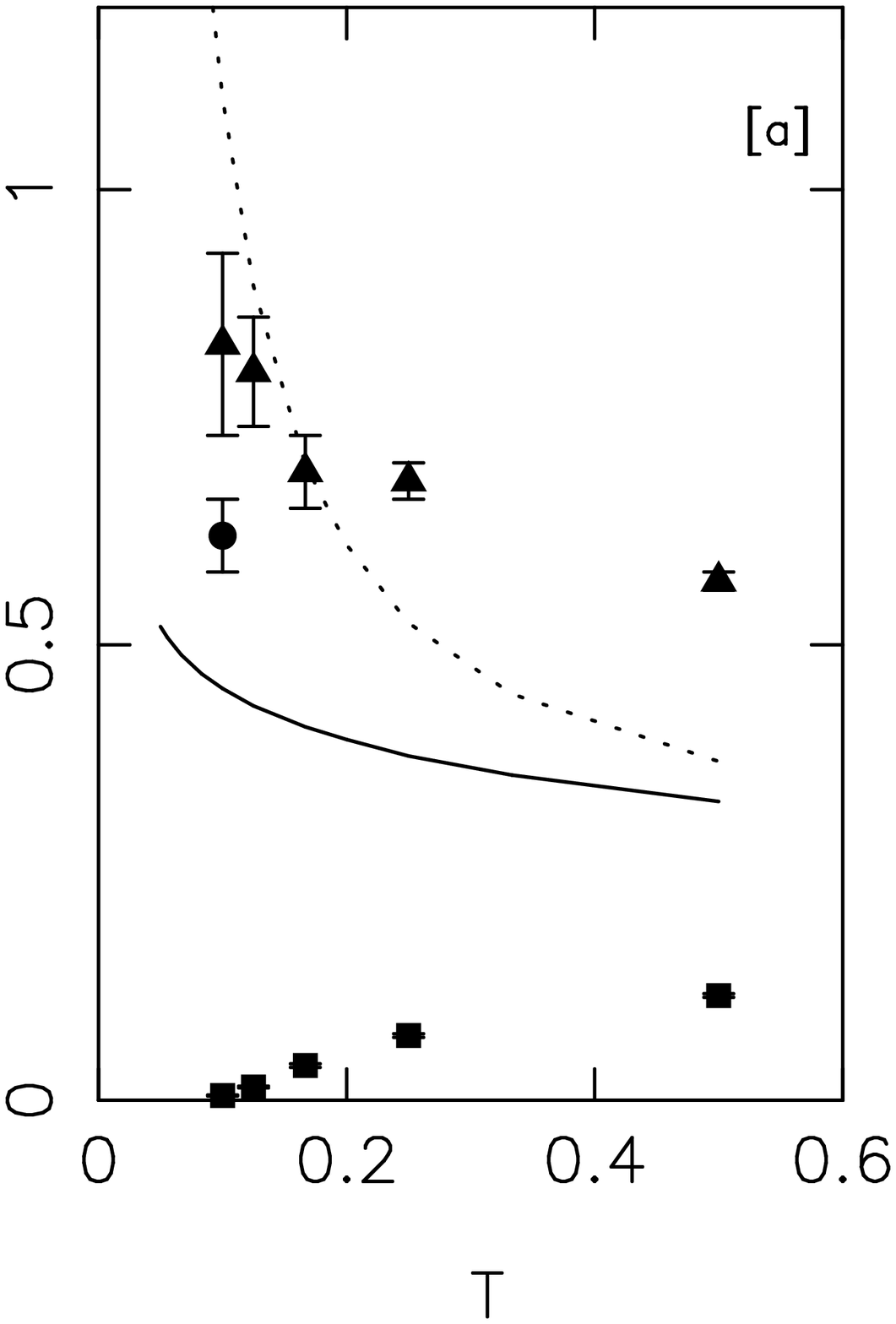,height=3.5in,width=4.in}
\vskip-13mm
\psfig{file=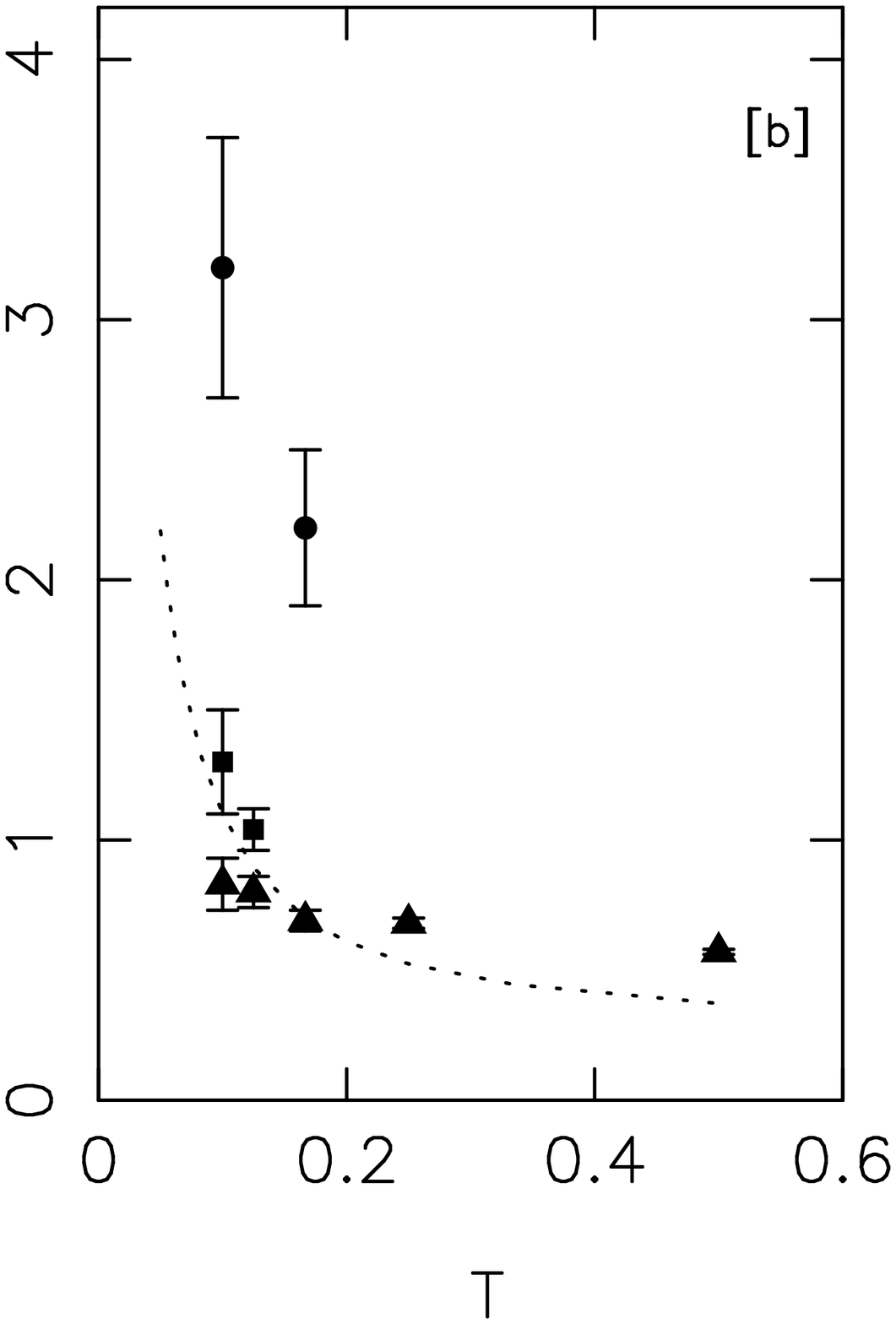,height=3.5in,width=4.in}
\vskip-3mm
\caption{Uniform susceptibility $\chi(0,0)$ vs.~$T$ on a $8\times 8$ lattice.
(a) $\chi(0,0)$ for $U=4$ and $\Delta=1.6$ (triangles); 
$\chi(0,0)$ for $U=4$, $T=t/10$, and 
$\Delta=0.0$, 0.4, and 0.8 (circle) are indistinguishable within 
the error-bars.
Also shown is the non-interacting $\kappa=\chi(0,0)$ 
for $\Delta=0.0$ (dotted line) and
$\Delta=1.6$ (full line). The compressibility $\kappa$ 
for $U=4$ and $\Delta=1.6$ 
(squares) vanishes exponentially at low temeratures.
The key point of this data is that while randomness
suppresses the susceptibility at $U=0$, it enhances it
for $U$ nonzero.
(b)  $\kappa=\chi(0,0)$ for $U=\Delta=0$ (dotted line)
and $\chi(0,0)$ for $U=4, \Delta=1.6$ (triangles) are again shown as in (a).
However, here they are now compared with 
$\chi(0,0)$ for $\Delta=2.4$ (squares) and for the
case of correlated hopping randomness with $\Delta'=1.6$ (circles) (see text).}
\end{figure}

\noindent
increases for lower temperatures
and is enhanced by almost one order of magnitude at $T=t/10$ compared to the 
case without disorder.
Although the statistical errors and limited temperatures do not allow
to extract the functional dependence on $T$, this dramatic increase
supports the existence of a finite density of localized moments. 

As in the interacting case without disorder, the compressibility, $\kappa$, 
shows for $\Delta=1.6$ activated behavior with temperature (Fig.~9a) with an 
estimated charge gap of 0.5 \cite{assaad}. A quite similar behavior is found 
even for stronger disorder $\Delta=2.4$ and also for the correlated hopping 
randomness with $\Delta'=1.6$ (not shown): $\kappa$ is very small 
$(\sim O(10^{-3}))$ and decreases strongly for lower $T$ consistent 
with an activated behavior.
Hence the systems remains uncompressible even in the absence of long
range magnetic order.
 
%
%
\section{Summary and Conclusions}

In this paper we have presented Quantum Monte Carlo
calculations for the magnetic phase diagram of
the disordered repulsive Hubbard model at half-filling.
When the randomness is in the site energy, the sign problem
precludes the study of large systems at low temperatures necessary
for a proper finite size scaling analysis.  However, we are able
to see on 4x4 lattices the destruction of local moments and longer
range spin-spin correlations.  We also studied the compressibility $\kappa$ 
as a function of disorder strength.  The incompressible state with
$\kappa=0$ at nonzero $U$
in the absence of disorder is replaced by a finite $\kappa$ 
at $\Delta \approx 2$, indicating a disorder induced closing of
the charge gap.

When the randomness is in the hoppings $t_{\bf ij}$ a particle-hole
transformation proves that the product of
determinants which gives the Boltzmann weight is positive at half-filling.
That is, there is no sign problem at arbitrarily low temperatures
and large lattices.  In this case we were able to do a finite
size scaling analysis of our results and determined $\Delta_{c} \approx 1.4$
for the destruction of antiferromagnetism.  This value was
in reasonable agreement with that obtained for the disordered
Heisenberg model \cite{sandvik2}.  The paramagnetic phase is driven
by singlet formation.  Unlike the random site energy case, the local
moments remain well-formed through the transition.
The strong enhancement of the uniform susceptibility at low temperatures,
in particular for the correlated hopping randomness, indicates the formation
of effectively localized magnetic moments.
The compressibilty is apparently activated even for strong disorder
where there are no long range magnetic correlations.

We are currently studying the dependence of the density of states in this
model as it has recently been suggested that this may provide an alternate
way to describe transitions in the Anderson-Hubbard model
and its experimental realizations.\cite{BELITZ2}

%
%

\section{Acknowledgments}
\medskip
R.~T.~S.~is supported by grant NSF-DMR-9528535.
M.~U.~is supported by a grant from the Office of Naval Research,
ONR N00014-93-1-0495 and by the Deutsche Forschungsgemeinschaft.
Computations were carried out at the San Diego Supercomputer Center.
\medskip

%
%

$^*$ e-mail: ulmke@physik.uni-augsburg.de

%
%
%
%
%



\begin{references}
\bibitem{finitelattices} 
Note, however that on finite lattices there is a Curie contribution 
to the compressibility ($\kappa\propto T^{-1}$) for temperatures smaller than
the smallest energy level distance $\propto 2\pi/N_x$. The weight of the
Curie divergence also vanishes like $\propto 1/N_x$.
\bibitem{percolation} 
Strictly at $J_2=0$ (dilute limit) the lattice is disconnected for 
concentrations $p$ below the percolation threshhold of $p_{perc}=0.5$.
In the limit of small $J_2$, however, one expects a finite (not vanishing) 
concentration of strong bonds to be necessary to destroy LRO. \cite{sandvik2} 
This discontinuity at $J_2=0$ is not described by the simple criterion
$v=v_c$ which provides a continous phase boundary up to the point
$(p=0,J_2=0)$. The disordered phase at $J_2=0$ extends to a concentration
$p_c=1-v_c$ which was found to be larger than $p_{perc}$ due to 
quantum fluctuations.
\end{references}
\end{document}